\documentclass[aps,preprint,amsmath]{revtex4}

\usepackage{graphicx}
\usepackage{pstricks}

\def\al{\alpha}
\def\be{\beta}

\def\de{\delta}
\def\ep{\epsilon}
\def\ve{\varepsilon}
\def\ze{\zeta}
\def\et{\eta}

\def\ka{\kappa}
\def\la{\lambda}

\def\rh{\rho}

\def\si{\sigma}

\def\om{\omega}

\def\cL{{\cal L}}

\def\cR{{\cal R}}
\def\cS{{\cal S}}
\def\cV{{\cal V}}

\def\mn{{\mu\nu}}

\def\kl{{\ka\la}}

\def\rs{{\rh\si}}

\newcounter{tc1}\newcounter{tc2}
\newcounter{tr1}\newcounter{tr2}
\def\half{{\textstyle{1\over 2}}}

\def\qrt{\tfrac14}

\def\tr{{\rm{tr}}}

\def\lsim{\mathrel{\rlap{\lower4pt\hbox{\hskip1pt$\sim$}}
    \raise1pt\hbox{$<$}}}
\def\gsim{\mathrel{\rlap{\lower4pt\hbox{\hskip1pt$\sim$}}
    \raise1pt\hbox{$>$}}}
\def\sqr#1#2{{\vcenter{\vbox{\hrule height.#2pt
         \hbox{\vrule width.#2pt height#1pt \kern#1pt
         \vrule width.#2pt}
         \hrule height.#2pt}}}}

\def\prt{\partial}

\def\etal{{\it et al.}}

\def\dc{\circ}

\newcommand{\beq}{\begin{equation}}
\newcommand{\eeq}{\end{equation}}
\newcommand{\bea}{\begin{eqnarray}}
\newcommand{\eea}{\end{eqnarray}}
\newcommand{\bit}{\begin{itemize}}
\newcommand{\eit}{\end{itemize}}
\newcommand{\rf}[1]{(\ref{#1})}

\newlength{\h}
\def\newtableau#1#2{\psset{unit=12pt,linewidth=0.5pt}%
  \setlength{\h}{#2\psunit}\setlength{\h}{0.5\h}\addtolength{\h}{-0.3\psunit}
  \begin{pspicture}[shift=-\h](#1,#2)\small%
    \setcounter{tc1}{0}\setcounter{tc2}{1}%
    \setcounter{tr1}{#2}\setcounter{tr2}{#2}\addtocounter{tr1}{-1}%
    \psline(0,0)(0,#2)(#1,#2)}
\def\endtableau{\end{pspicture}}
\def\newbox#1{%
  \psline(\value{tc1},\value{tr1})(\value{tc2},\value{tr1})(\value{tc2},\value{tr2})%
  \rput(\value{tc1},\value{tr1}){\rput(0.5,0.5){#1}}
  \addtocounter{tc1}{1}\addtocounter{tc2}{1}}
\def\longbox#1{%
  \addtocounter{tc2}{1}%
  \psline(\value{tc1},\value{tr1})(\value{tc2},\value{tr1})(\value{tc2},\value{tr2})%
  \rput(\value{tc1},\value{tr1}){\rput(1.0,0.5){#1}}
  \addtocounter{tc1}{2}\addtocounter{tc2}{1}}
\def\newrow{%
  \addtocounter{tr1}{-1}\addtocounter{tr2}{-1}%
  \setcounter{tc1}{0}\setcounter{tc2}{1}}

\def\bc#1#2{\left(\begin{smallmatrix} #1 \\ #2 \end{smallmatrix}\right)}

\def\cd#1{{s}^{(#1)}}
\def\bd#1{{q}^{(#1)}}
\def\dd#1{{k}^{(#1)}}

\def\cK{\mathcal K}

\def\cKd{{\cK}^{(d)}{}}
\def\cKHatd{\widehat{\cK}^{(d)}{}}

\def\S{s}

\def\Sbz{{\S}^{(d,1)}}
\def\Sba{{\S}^{(d,2)}}

\def\Q{q}

\def\Qaz{{\Q}^{(d,1)}}
\def\Qbz{{\Q}^{(d,2)}}
\def\Qaa{{\Q}^{(d,3)}}
\def\Qcz{{\Q}^{(d,4)}}
\def\Qba{{\Q}^{(d,5)}}

\def\K{k}

\def\Kzz{{\K}^{(d,1)}}
\def\Kaz{{\K}^{(d,2)}}
\def\Kbz{{\K}^{(d,3)}}
\def\Kcz{{\K}^{(d,4)}}

\def\s{s}
\def\st{\bar\s}
\def\std#1{{\st}^{(#1)}}

\def\dual#1{*{#1}}
\def\zdual#1{\star{#1}}

\def\MB{\breve{M}}
\def\TR{\varpi}

\def\zB{{\widehat z}}

\def\sB{{\widehat s}}

\def\zU{{z'}}
\def\rU{{r'}}
\def\sU{{s'}}
\def\zeU{{\ze'}}

\def\zUdual#1{{\star'}{#1}}

\def\cSB{{\widehat\cS}}

\begin{document}

\title{
Lorentz and Diffeomorphism Violations in Linearized Gravity
}

\author{V.\ Alan Kosteleck\'y$^1$ and Matthew Mewes$^2$}

\affiliation{$^1$Physics Department, Indiana University,
Bloomington, Indiana 47405, USA\\
$^2$Physics Department, California Polytechnic State University, 
San Luis Obispo, California 93407, USA}

\date{IUHET 626, December 2017}

\begin{abstract}

Lorentz and diffeomorphism violations are studied in linearized gravity 
using effective field theory.
A classification of all gauge-invariant and gauge-violating terms is given.
The exact covariant dispersion relation for gravitational modes 
involving operators of arbitrary mass dimension is constructed,
and various special limits are discussed.

\end{abstract}

\maketitle

The foundational symmetries of General Relativity (GR)
include diffeomorphisms and local Lorentz transformations.
The former act on the spacetime manifold, 
while the latter act in the tangent space.
These two types of transformations are partially linked through the vierbein,
which provides a tool for moving objects 
between the manifold and the tangent space.
The proposal that Lorentz invariance
might be broken in an underlying theory 
of gravity and quantum physics such as strings 
\cite{ksp,reviews}
naturally raises various questions about the relationship 
between diffeomorphism violation and Lorentz violation
and about the associated phenomenological signals.
These questions can be studied independently of specific models
using gravitational effective field theory
\cite{ak04}.
Here,
following a brief summary of the current status and results,
we develop a model-independent framework for studying these issues 
in linearized gravity.
This limit  
provides a comparatively simple arena for exploration,
and it is crucial for experimental analyses of gravitational waves 
and of gravitation in the Newton and post-Newton limits.

A generic treatment of Lorentz violation in Minkowski spacetime 
in the absence of gravity is comparatively straightforward
using effective field theory 
\cite{ck}.
In this context,
the role of diffeomorphisms and local Lorentz transformations
is played by translations and Lorentz transformations that act globally
and combine to form the Poincar\'e group.
The two symmetries can be broken independently,
and a physical breaking of either one 
can be represented in terms of nonzero background fields
in an effective field theory.
The breaking of either can be spontaneous or explicit.
Spontaneous breaking occurs when the background is dynamical,
which means that it must satisfy the equations of motion 
and that it comes with fluctuations in the form of Nambu-Goldstone modes 
\cite{ng}
and possibly also massive modes.
In most applications of spontaneous breaking,
the background satisfies the equations of motion in vacuum
and can therefore be viewed as the vacuum expectation value.
In contrast,
explicit breaking is a consequence of a prescribed background,
which is typically off shell
and has no associated fluctuations.
Much of the phenomenological literature 
investigating Lorentz violation in Minkowski spacetime
assumes for simplicity that global spacetime translations are preserved
in an approximately local inertial frame,
canonically taken to be the Sun-centered frame
\cite{sunframe}. 
This guarantees conservation of energy and momentum,
so phenomenological signals are restricted
to violations of the conservation laws for generalized angular momenta. 
A large body of experimental studies constrains 
this type of Lorentz violation
\cite{tables}.

In the presence of gravity,
the situation becomes more involved.
One complication arises because
diffeomorphisms and local Lorentz transformations
act on objects in different spaces 
that can be linked via the vierbein,
which can relate the corresponding violations.
In the case of spontaneous breaking,
for example,
the vacuum expectation values are on shell
and a nonzero background on the spacetime manifold
implies one in the tangent space and vice versa.
As a result,
diffeomorphism violation occurs if and only if local Lorentz violation does
\cite{bk05}.
More intuitively,
local Lorentz violation can be understood 
as a background direction dependence in a local freely falling frame
\cite{ak04}.
Transporting this to the spacetime manifold via the vierbein
then guarantees the existence 
of a direction dependence on the spacetime manifold
and hence diffeomorphism violation.

Another complication for gravity concerns conservation laws
and arises from the difference between spontaneous and explicit breaking. 
In general,
a theory invariant under local transformations 
comes with covariantly conserved currents
\cite{noether}.
In spontaneous breaking,
the full theory remains invariant under the transformations
and the symmetry is only hidden
\cite{coleman}.
The currents remain conserved
even though the background is unchanged by the transformations 
because the background fluctuations 
transform in a nonstandard way to compensate.
This contrasts with explicit breaking,
when the current conservation laws fail to hold.

In GR,
local Lorentz invariance implies symmetry of the energy-momentum tensor
while diffeomorphism invariance implies its covariant conservation
\cite{mtw}.
In theories with spontaneous diffeomorphism and local Lorentz violation,
these current-conservation laws are unaffected:
an energy-momentum tensor for the full theory remains covariantly conserved
and it is always possible to make it symmetric 
\cite{ak04}.
However,
if explicit breaking occurs,
then there is no guarantee that the energy-momentum tensor 
is explicitly conserved or symmetric,
and as a result a theory with explicit breaking can be inconsistent
or require reformulation within Finsler geometry 
\cite{ak04,ak11}.
For sufficiently involved models,
this situation can be rescued by the additional modes 
that appear in theories with explicit diffeomorphism 
and local Lorentz violation
\cite{rb}.
These additional modes
arise because in explicit breaking it becomes impossible
to remove all four diffeomorphism degrees of freedom 
and six local Lorentz degrees of freedom from the vierbein.
In some models,
these additional modes can be constrained to restore 
the covariant conservation and symmetry of the energy-momentum tensor.
The additional modes are the counterparts in explicit breaking
of the Nambu-Goldstone modes appearing in spontaneous breaking.
Indeed,
they can be understood as Nambu-Goldstone excitations
of Stueckelberg fields
\cite{stueckelberg,ags}. 

The above results have several implications 
for the phenomenology of diffeomorphism and local Lorentz violations
in gravity.
If the breaking is explicit,
the challenge lies in establishing the consistency of theory
and,
if achieved,
then in determining the effects of the additional modes
on observational signals.
In contrast,
if the breaking is spontaneous,
the Nambu-Goldstone and massive fluctuations 
can play the role of new forces affecting the phenomenology 
and so must be taken into account in analyzing experimental signals.
Model-independent techniques for this have been developed 
both in the pure-gravity and in the matter-gravity sectors 
\cite{bk06,kt09,se09,al10,kt11,bt11,jt12,bkx14,yb15,je15,bh17}
and applied to obtain model-independent constraints 
on diffeomorphism and local Lorentz violation in gravity
from a variety of experimental tests 
\cite{tables,2007Battat,2007MullerInterf,2009Chung,%
2010Bennett,2010Panjwani,2010Altschul,2011Hohensee,%
2012Iorio,2013Bailey,2014Shao,he15,%
le16,bo16,lk,hust15,hustiu,hust16,kt15,jt16,km16,qb16,tz16,%
yu16,ms16,2017Abbott,km17,2017Chen,2016Flowers,2017Shao,2017Wang}.

An alternative model-independent approach
to studing both spontaneous and explicit 
diffeomorphism and local Lorentz violation
uses linearized effective field theory for gravity,
formulated to incorporate gauge and Lorentz violation
\cite{km16}.
In this context,
gauge transformations are linearized diffeomorphisms 
of the metric fluctuation.
This technique yields an explicit construction and classification 
of the general quadratic Lagrange density in effective field theory
with gauge invariance at linearized level.
It also permits construction of the general covariant dispersion relation 
and investigation of the properties of the corresponding gravitational modes.
These results have been applied to obtain model-independent constraints
on linearized coefficients for Lorentz violation
using gravitational waves 
\cite{km16,2017Abbott}
and tests of gravity at short range  
\cite{km17,2017Chen}.
In the present work,
we extend this approach to explicit gauge breaking.
We construct and classify all possible terms
for the quadratic Lagrange density in gravitational effective field theory
with explicit gauge violation,
and we derive the corresponding covariant dispersion relation
required for experimental applications.

To perform the linearization, 
we expand the dynamical metric $g_\mn$ in a flat-spacetime background 
with Minkowski metric,
$g_\mn = \et_\mn+ h_\mn$.
A generic term of mass dimension $d\geq 2$
in the Lagrange density
for the linearized gravitational effective field theory
can then be written as
\beq
\cL_{\cKd} = \qrt h_\mn \cKHatd^{\mn\rs} h_\rs ,
\label{L1}
\eeq
where $\cKHatd^{\mn\rs}$ is the product of a coefficient
$\cKd^{\mn\rs\ve_1\ve_2\ldots\ve_{d-2}}$
with $d-2$ derivatives
$\prt_{\ve_1} \prt_{\ve_2}\ldots \prt_{\ve_{d-2}}$.
The coefficients $\cKd^{\mn\rs\ve_1\ve_2\ldots\ve_{d-2}}$
have mass dimension $4-d$ 
and are assumed constant and small.
The complete traces of these coefficients
control Lorentz-invariant terms in  $\cL_{\cKd}$,
while the other components govern Lorentz violation.
To contribute nontrivially to the equations of motion,
the operator $\cKHatd^{\mn\rs}$ must satisfy the requirement
$\cKHatd^{(\mu\nu)(\rh\si)} \neq \pm \cKHatd^{(\rh\si)(\mu\nu)}$,
where the upper sign holds for odd $d$
and the lower one for even $d$.

The action is invariant
under the usual gauge transformations 
$h_\mn \to h_\mn + \prt_\mu \xi_\nu + \prt_\nu \xi_\mu$
when the condition 
$\cKHatd^{(\mu\nu)(\rh\si)} \prt_{\nu}
= \pm \cKHatd^{(\rh\si)(\mu\nu)} \prt_{\nu}$
holds.
Assuming this condition,
the operators $\cKHatd^{\mn\rs}$ can be constructed explicitly,
using standard methods in group theory
\cite{group}.
They are found to span three representation classes 
\cite{km16}.
For the present work,
we have extended this construction
by relaxing the requirement of gauge invariance.
Decomposing the operator $\cKHatd^{\mn\rs}$
into irreducible pieces then yields another 11 representation classes.
This shows that a total of only 14 independent classes of operators 
can appear in any linearized gravitational effective field theory,
whether or not the Lorentz and gauge invariances hold.
These 14 classes therefore characterize
all possible phenomenological effects in linearized gravity,
including effects on the propagation of gravitational waves
and in the Newton and post-Newton limits. 

To simplify the notation in what follows,
we denote indices contracted into a derivative 
as a circle index $\dc$,
with $n$-fold contractions denoted as $\dc^n$.
With this convention,
the generic operator $\cKHatd^{\mn\rs}$
can be written as
$\cKHatd^{\mn\rs}=\cKd^{\mn\rs{\dc^{d-2}}}$.
Also,
we denote the 14 representation classes
as indicated in the first column of Table \ref{table1}.
To obtain the term in the Lagrange density \rf{L1} 
associated to a given class,
it suffices to replace $\cKHatd^{\mn\rs}$
with the operator listed.
The second column displays the index symmetries of each class 
using Young tableaux.
The Table also lists some properties of each class.
The third column indicates whether the operator is fully gauge invariant,
and the fourth column displays the handedness under CPT 
of the associated term in the Lagrange density. 
Each class can occur only for even or for odd $d$
and for $d$ above a minimal value,
as shown in the next column.
The final column
lists the total number of independent components
appearing in the coefficient $\cKd^{\mn\rs\ve_1\ve_2\ldots\ve_{d-2}}$
for fixed $d$.

\begin{table*}
\def\vsp{&&&&\\[-8pt]}
\tabcolsep10pt
\begin{tabular}{l|c|c|c|c|c}
	&		&	Gauge	&		&		&		\\	
\multicolumn{1}{c|}{Operator $\cKHatd^{\mn\rs}$} 	&	Tableau	&	invariant	&	CPT	&	$d$	&	Number	\\	
		\hline\hline										\vsp
$\cd{d}{}^{\mu\rh\dc\nu\si\dc\dc^{d-4}}$	&	$\newtableau{4}{3} \newbox{$\mu$}\newbox{$\nu$}\longbox{$\cdots$} \newrow\newbox{$\rh$}\newbox{$\si$} \newrow\newbox{$\dc$}\newbox{$\dc$} \endtableau$	&	yes	&	even	&	even, $\geq 4$	&	$(d-3)(d-2)(d+1)$	\\	\vsp
$\Sbz{}^{\mu\rh\nu\si\dc^{d-2}}$	&	$\newtableau{4}{2} \newbox{$\mu$}\newbox{$\nu$}\longbox{$\cdots$} \newrow\newbox{$\rh$}\newbox{$\si$} \endtableau$	&	no	&	even	&	even, $\geq 2$	&	$(d-1)(d+2)(d+3)$	\\	\vsp
$\Sba{}^{\mu\rh\dc\nu\si\dc\dc^{d-4}}$	&	$\newtableau{5}{3} \newbox{$\mu$}\newbox{$\nu$}\newbox{$\dc$}\longbox{$\cdots$} \newrow\newbox{$\rh$}\newbox{$\si$} \newrow\newbox{$\dc$} \endtableau$	&	no	&	even	&	even, $\geq 4$	&	$\tfrac43(d-2)d(d+2)$	\\	\vsp
		\hline										\vsp
$\bd{d}{}^{\mu\rh\dc\nu\dc\si\dc\dc^{d-5}}$	&	$\newtableau{5}{3} \newbox{$\mu$}\newbox{$\nu$}\newbox{$\si$}\longbox{$\cdots$} \newrow\newbox{$\rh$}\newbox{$\dc$}\newbox{$\dc$} \newrow\newbox{$\dc$} \endtableau$	&	yes	&	odd	&	odd, $\geq 5$	&	$\tfrac52(d-4)(d-1)(d+1)$	\\	\vsp
$\Qaz{}^{\mu\rh\nu\si\dc\dc^{d-3}}$	&	$\newtableau{6}{2} \newbox{$\mu$}\newbox{$\nu$}\newbox{$\si$}\newbox{$\dc$} \longbox{$\cdots$} \newrow\newbox{$\rh$} \endtableau$	&	no	&	odd	&	odd, $\geq 3$ 	&	$\tfrac12(d+1)(d+3)(d+4)$ 	\\	\vsp
$\Qbz{}^{\mu\rh\nu\dc\si\dc^{d-3}}$	&	$\newtableau{5}{2} \newbox{$\mu$}\newbox{$\nu$}\newbox{$\si$}\longbox{$\cdots$} \newrow\newbox{$\rh$}\newbox{$\dc$} \endtableau$	&	no	&	odd	&	odd, $\geq 3$ 	&	$(d-1)(d+2)(d+3)$ 	\\	\vsp
$\Qaa{}^{\mu\rh\dc\nu\si\dc^{d-3}}$	&	$\newtableau{5}{3} \newbox{$\mu$}\newbox{$\nu$}\newbox{$\si$}\longbox{$\cdots$} \newrow\newbox{$\rh$} \newrow\newbox{$\dc$} \endtableau$	&	no	&	odd	&	odd, $\geq 3$	&	$\half d(d+1)(d+3)$	\\	\vsp
$\Qcz{}^{\mu\rh\nu\dc\si\dc\dc\dc^{d-5}}$	&	$\newtableau{6}{2} \newbox{$\mu$}\newbox{$\nu$}\newbox{$\si$}\newbox{$\dc$}\longbox{$\cdots$} \newrow\newbox{$\rh$}\newbox{$\dc$}\newbox{$\dc$} \endtableau$	&	no	&	odd	&	odd, $\geq 5$	&	$\tfrac53(d-3)(d+1)(d+2)$	\\	\vsp
$\Qba{}^{\mu\rh\dc\nu\dc\si\dc\dc^{d-5}}$	&	$\newtableau{6}{3} \newbox{$\mu$}\newbox{$\nu$}\newbox{$\si$}\newbox{$\dc$}\longbox{$\cdots$} \newrow\newbox{$\rh$}\newbox{$\dc$} \newrow\newbox{$\dc$} \endtableau$	&	no	&	odd	&	odd, $\geq 5$	&	$\tfrac43(d-2)d(d+2)$ 	\\	\vsp
		\hline										\vsp
$\dd{d}{}^{\mu\dc\nu\dc\rh\dc\si\dc\dc^{d-6}}$	&	$\newtableau{6}{2} \newbox{$\mu$}\newbox{$\nu$}\newbox{$\rh$}\newbox{$\si$}\longbox{$\cdots$} \newrow\newbox{$\dc$}\newbox{$\dc$}\newbox{$\dc$}\newbox{$\dc$} \endtableau$	&	yes	&	even	&	even, $\geq 6$	&	$\tfrac52(d-5)d(d+1)$	\\	\vsp
$\Kzz{}^{\mn\rs\dc^{d-2}}$	&	$\newtableau{6}{1} \newbox{$\mu$}\newbox{$\nu$}\newbox{$\rh$}\newbox{$\si$}\longbox{$\cdots$} \endtableau$	&	no	&	even	&	even, $\geq 2$	&	$\tfrac16(d+3)(d+4)(d+5)$	\\	\vsp
$\Kaz{}^{\mu\dc\nu\rs\dc\dc^{d-4}}$	&	$\newtableau{7}{2} \newbox{$\mu$}\newbox{$\nu$}\newbox{$\rh$}\newbox{$\si$}\newbox{$\dc$}\longbox{$\cdots$} \newrow\newbox{$\dc$} \endtableau$	&	no	&	even	&	even, $\geq 4$	&	$\tfrac12(d+1)(d+3)(d+4)$	\\	\vsp
$\Kbz{}^{\mu\dc\nu\dc\rh\si\dc^{d-4}}$	&	$\newtableau{6}{2} \newbox{$\mu$}\newbox{$\nu$}\newbox{$\rh$}\newbox{$\si$}\longbox{$\cdots$} \newrow\newbox{$\dc$}\newbox{$\dc$} \endtableau$	&	no	&	even	&	even, $\geq 4$	&	$(d-1)(d+2)(d+3)$	\\	\vsp
$\Kcz{}^{\mu\dc\nu\dc\rh\dc\si\dc\dc^{d-6}}$	&	$\newtableau{7}{2} \newbox{$\mu$}\newbox{$\nu$}\newbox{$\rh$}\newbox{$\si$}\newbox{$\dc$}\longbox{$\cdots$} \newrow\newbox{$\dc$}\newbox{$\dc$}\newbox{$\dc$} \endtableau$	&	no	&	even	&	even, $\geq 6$	&	$\tfrac53(d-3)(d+1)(d+2)$	 	
\end{tabular}
\caption{
\label{table1}
Operators in the quadratic action for linearized gravity.
}
\end{table*}

The quadratic approximation $\cL_0$
to the Lagrange density for the Einstein-Hilbert action 
can conveniently be written in the form 
\bea
\cL_{0} &=& 
\qrt \ep^{\mu\rh\al\ka} \ep^{\nu\si\be\la} 
\et_\kl h_\mn \prt_\al\prt_\be h_\rs.
\eea
This gauge- and Lorentz-invariant term is constructed
from a piece of the coefficient $\cd{4}{}^{\mu\rh\al\nu\si\be}$
in the first line of Table \ref{table1}.
The complete Lagrange density incorporating all the operators
in Table \ref{table1}
can then be expressed as 
\bea
\cL &=& \cL_0 + 
\qrt h_\mn 
\sum_{\cK,d} 
\cKHatd^{\mn\rs} h_\rs ,
\label{L2}
\eea
where the sum is over all the representation classes $\cKHatd^{\mn\rs}$ 
shown in Table \ref{table1}
and also over all allowed dimensions $d$ for each class. 
Larger values of $d$ introduce higher powers of momenta 
and so the corresponding terms in the effective field theory
are expected to be more suppressed.
In practice,
to avoid possible issues with interpretation of the infinite sum of terms,
the sum over $d$ can be truncated at some value 
or restricted to specific choices of $d$. 

The equations of motion for the metric fluctuation $h_{\mu\nu}$ 
can be found from the Lagrange density.
Performing Fourier transforms to convert to momentum space,
where $\prt_\mu \to ip_\mu$,
the equations of motion can be written in the form
\beq
M_{\ka\la}{}^{\mu\nu} h_{\mu\nu} = 0.
\label{eqmot}
\eeq
The operator $M_{\ka\la}{}^{\mu\nu}= M_{\ka\la}{}^{\mu\nu}(p)$
can be understood as a square $10\times10$ matrix
that acts on a 10-component vector $h_{\mn}$.
If no gauge invariances are present,
the dispersion relation for the gravitational modes is obtained 
by setting the determinant of the matrix to zero. 
However,
in the presence of partial or full gauge invariance,
finding a dispersion relation for the physical modes is more complicated 
because $M_{\ka\la}{}^{\mu\nu}$ contains a null space. 
Nonetheless,
a covariant dispersion relation can be found
using methods from exterior algebra,
as we show next.
The technique presented here
is a generalization of the method developed by us
for the study of the photon sector of the SME
\cite{km09} 
and independently by Itin 
for studies of premetric electrodynamics
\cite{itin}.

The key idea is to treat $h_{\mn}$ 
as an element of a 10-dimensional complex vector space
and $M_{\ka\la}{}^{\mu\nu}$ as a linear map on the space.
To keep the discussion general,
we work with an $N$-dimensional complex vector space $\cV$
and the exterior algebra $\wedge\cV$ over $\cV$.
In terms of an arbitrary set of basis vectors
$\{v^a\}$, $a = 1,2,\ldots,N$,
we write an $n$-vector $\om$ as
\beq
\om = \tfrac{1}{n!} 
\om_{a_1a_2\ldots a_n} v^{a_1}\wedge v^{a_2}\wedge \cdots \wedge v^{a_n} ,
\eeq
and take its Hodge dual $\dual\om$
to have components given by
\beq
(\dual\om)^{a_1 \ldots a_{N-n}} = 
\tfrac{1}{n!} \ep^{a_1\ldots a_{N-n}b_1\ldots b_n} \om_{b_1\ldots b_n} .
\label{hodge}
\eeq
Given a linear map $M:\cV\to\cV$
taking an arbitrary vector $x\in\cV$ to a vector $y=M\cdot x$,
we can construct a natural linear map $\wedge^nM$ between $n$-vectors.
For $n$ arbitrary vectors $\{x^1,x^2,\ldots, x^n\}$,
we define 
\beq
y^1\wedge y^2 \wedge \cdots \wedge y^n
= \wedge^n M
(x^1\wedge x^2 \wedge \cdots \wedge x^n) .
\eeq
In components this gives 
\beq
{(\wedge^n M)_{a_1a_2\ldots a_n}}^{b_1b_2\ldots b_n}
= \tfrac{1}{n!} {M_{[a_1}}^{b_1}{M_{a_2}}^{b_2}\ldots{M_{a_n]}}^{b_n} .
\eeq

Let $r$ be the rank of $M$
and let $s$ be the dimension of the null space,
so $r+s=N$.
Denote the null space of $M$ by $\cS$ and its complement by $\cR$,
so that $\cV = \cR \oplus \cS$.
Let $\{z^1,\ldots,z^s\}$ be a set of vectors spanning $\cS$,
and let $\{x^1,\ldots,x^r,z^1,\ldots,z^s\}$
span $\cV$.
Any $n$-vector can then be expressed as a linear
combinations of wedge products of $n$ of these vectors.
This implies that $\wedge^n M = 0$ for $r<n$.
Also,
the rank of $\wedge^n M$ for $n\leq r$ is $\bc{r}{n}$ 
because the dimension of the image of $\wedge^n M$
matches the number of $n$-vectors 
constructed from the vectors $\{x^1,\ldots,x^r\}$.
In particular,
while the map $M$ is rank $r$,
the map $\wedge^r M$ is rank 1.

The dual map $\dual\wedge^r M$ can be constructed 
using the Hodge dual \rf{hodge}.
However, 
both $\wedge^r M$ and $\dual\wedge^r M$ 
incorporate a null space,
which complicates the derivation of the dispersion relation. 
To account explicitly for the null space of $\wedge^n M$,
we can work instead with a modified dual $\zdual{\wedge^n M}$.
Introducing $\ze = \dual(z^1\wedge \cdots \wedge z^s)$,
some consideration reveals that we can write 
\bea
(\wedge^n M)_{a_1\ldots a_n}{}^{b_1\ldots b_n}
&=& 
\tfrac{1}{(r-n)!}\tfrac{1}{(r-n)!}
\ze^*_{a_1\ldots a_n c_1\ldots c_{r-n}}
\ze^{b_1\ldots b_n d_1\ldots d_{r-n}}
(\zdual{\wedge^n M})_{d_1\ldots d_{r-n}}{}^{c_1\ldots c_{r-n}} .
\label{zdual}
\eea
In the special case where $n=r$,
we see that $\zdual{\wedge^r M}$ is a scalar obeying 
$\wedge^r M = \zdual{\wedge^r M} \, \ze^* \otimes \ze$,
which implies
\beq
\wedge^r M = \tr(\wedge^r M) \, \frac{\ze^* \otimes \ze}{\ze^*\cdot\ze} .
\eeq
The inverse relation for the modified dual
can be obtained from Eq.\ \rf{zdual}.
After some manipulation, 
we find
\bea
(\zdual{\wedge^n M})_{b_1\ldots b_{r-n}}{}^{a_1\ldots a_{r-n}}
&=& 
\left(\tfrac{r!}{n!\, \ze^*\cdot\ze}\right)^2
\ze^{a_1\ldots a_{r-n} c_1\ldots c_n}
\ze^*_{b_1\ldots b_{r-n} d_1\ldots d_n}
{(\wedge^n M)_{c_1\ldots c_n}}^{d_1\ldots d_n}.
\qquad
\label{moddual}
\eea
For the case $\cS = \emptyset$ so that $\cV = \cR$, 
the modified dual $\zdual{\wedge^n M}$
reduces to the usual dual $\dual{\wedge^n M}$.
Taking instead $r=n$ yields the modified scalar dual
\beq
\zdual{\wedge^r M} =
(\ze^\dual\cdot\ze)^{-1}\tr(\wedge^r M)
= |\ze\cdot\ze|^{-2}\ze^* \cdot \wedge^r M \cdot \ze .
\eeq
Note that since the modified scalar dual $\zdual{\wedge^r M}$
contains inverse powers of $\ze^*\cdot\ze$,
one might naively expect a singularity when $\ze^*\cdot\ze=0$.
However, 
continuity and the relation 
$\wedge^r M = \zdual{\wedge^r M} \, \ze^* \otimes \ze$
guarantee that
$\zdual{\wedge^r M}$ remains finite for nonzero $\ze$.
Similarly,
although $\zdual{\wedge^n M}$ may contain divergences,
they cannot contribute to $\wedge^n M$ in Eq.\ \rf{zdual}.

Our goal is to obtain the exact covariant dispersion relation
from equations of motion of the form $M\cdot x = 0$,
where $M$ depends on the momentum $p^\mu$,
while allowing for a possible nontrivial null space of $M$.
Typically,
we are interested in situations where 
it is convenient to split $M$ into two pieces,
$M = M_0 + \de M$.
This is useful,
for instance,
when $M_0$ is a standard expression 
or when calculations with $M_0$ can be performed in closed form.
In many applications the Lorentz violation can be taken small,
in which case the Lorentz-violating terms
can be placed in $\de M$ and treated perturbatively.
Note that the null spaces of $M$, $M_0$, and $\de M$ may all differ,
but the null space of any one must contain the intersection
of the null spaces of the other two.

To fix notation,
suppose $M_0$ has rank $r$ and $M$ has rank $\rU$.
Let the null space $\cS$ of $M_0$ 
be spanned by vectors $\{z^1,\ldots,z^s\}$,
and let the null space $\cS'$ of $M$ 
be spanned by vectors $\{\zU{}^1,\ldots,\zU{}^\sU\}$.
Define 
$\ze = \dual(z^1\wedge \cdots \wedge z^s)$
and 
$\zeU = \dual(\zU{}^1\wedge \cdots \wedge \zU{}^\sU)$,
and let 
$\zdual{\wedge^n M}$ be the modified dual \rf{moddual} constructed with $\ze$
and 
$\zUdual{\wedge^n M}$ be the modified dual constructed with $\zeU$.
Vectors in the null space $\cS$ 
represent trivial pure-gauge solutions of $M_0 \cdot x = 0$,
while vectors in $\cS'$
represent trivial solutions of $M \cdot x = 0$.

We seek nontrivial solutions to $M\cdot x =0$,
which exist if we can find $p_\mu$ 
that reduce the rank of $M$ by at least one.
The exact covariant dispersion relation 
arising from $M\cdot x =0$ 
can therefore be expressed as 
\beq
\zUdual{\wedge^{\rU} M} = 0.
\label{drfull}
\eeq
After expanding $\zUdual{\wedge^{\rU} M}$ in terms of $M_0$ and $\de M$,
some calculation reveals that this equation can be written as 
\beq
\sum_n 
\frac{n!}{\rU!\,(\rU-n)!}
\tr\left[(\zUdual{\wedge^{n}M_0}) 
\cdot (\wedge^{(\rU-n)} \de M)\right] =0 .
\label{dr1}
\eeq
The sum in this expression is understood to be 
limited to nonnegative wedge powers
and restricted by the ranks of $M_0$ and $\de M$.
For example, 
$n\leq r$ because larger values of $n$ 
produce $\wedge^{n}M_0 = 0$ and so cannot contribute.

The expression \rf{dr1} for the covariant dispersion relation 
is general and exact,
and it is convenient for applications 
where the unbroken gauge vectors $\{\zU{}^1,\ldots,\zU{}^\sU\}$
spanning $S'$ and thus the form of $\zeU$ are known.
However,
in some scenarios it is easier to work
with the broken gauge vectors instead. 
In particular,
in many situations of interest
the null space $\cS$ of $M_0$ contains the null space $\cS'$ of $M$,
$\cS\supseteq \cS'$,
so that $\de M$ acts to remove a subset of null vectors in $\cS$.
Then,
the ranks $r$ of $M_0$ and $\rU$ of $M$ satisfy $r\leq \rU$,
and the coranks satisfy $\sU\leq s$.
Null vectors in $\cS$ can then be split into those spanning $\cS'$ 
and those spanning the complement $\cSB = \cS - \cS'$.
Choosing a canonical ordering for definiteness,
we write 
$\{z^1,\ldots,z^s\} =
\{\zB^1,\ldots,\zB^\sB,\zU^1,\ldots,\zU{}^\sU\}$,
where $\sB = s-\sU = \rU-r$
is the number of gauge symmetries in $M_0$
that are broken in $M$.
Introducing
$\xi = (\zB^1\otimes \zB^{1*})\wedge\cdots\wedge (\zB^\sB\otimes\zB^{\sB*})$,
calculation then yields
an alternative form for the dispersion relation,
\beq
\sum_n \frac{n!(\rU-n)!}{\rU!\,(r-n)!^2}
\tr\left[\big((\zdual{\wedge^{n}M_0})\wedge\xi\big) 
\cdot (\wedge^{(\rU-n)} \de M)\right] =0 ,
\label{dr2}
\eeq
which holds for $\cS\supseteq \cS'$
and is convenient when the form of the broken gauge vectors
$\{\zB^1,\ldots,\zB^\sB\}$ is known.

A comparatively simple special case of the above 
arises when $M$ and $M_0$ have the same null space $\cS$,
so that $\ze=\zeU$.
In this case, 
$\de M$ preserves the gauge invariance of $M_0$
and the null space of $\de M$ must contain $\cS$.
The exact covariant dispersion relation then reduces to 
\beq
\sum_n \frac{n!}{r!\,(r-n)!}
\tr\left[(\zdual{\wedge^{n}M_0}) 
\cdot (\wedge^{(r-n)} \de M)\right] = 0 .
\label{dr3}
\eeq
For example,
with $M_0$ corresponding to the usual linearized Einstein-Hilbert term 
in the Lagrange density and taking $r=6$,
Eq.\ \rf{dr3} provides the exact covariant dispersion relation
for the general linearized gravitational theory 
formed by extending linearized GR with arbitrary gauge-invariant terms.
As another example,
if neither $M$ nor $M_0$ has any gauge symmetry then $r=10$,
and Eq.\ \rf{dr3} provides the covariant dispersion relation
for any $\de M$ with or without gauge invariance.

As an application of the above results,
consider linearized GR.
The linearized Einstein field equations take the form
$M_0{}_{\mn}{}^{\rs} h_\rs=0$.
The operator $M_0{}_{\mn}{}^{\rs}$ can be expressed as
\beq
M_0{}_{\mn}{}^{\rs}
= \half p^2 (\pi_{\mn}{}^{\rs} - P_\mn P^\rs) ,
\label{GRM}
\eeq
where the projections $P^\mn$ and $\pi_{\mn}{}^{\rs}$ given by
\beq
P^\mn = \et^\mn - \frac{p^\mu p^\nu}{p^2} ,
\quad
\pi_\mn{}^\rs = \half P_\mu{}^{(\rh}P_\nu{}^{\si)}
\eeq
are a subset of the standard spin-2 projection operators
\cite{proj1,proj2}.
Note that the trace of $P$ is 3,
while that of $\pi$ is 6.
The wedge products are found to be
\beq
\wedge^n M_0
= \frac{p^{2n}}{2^n} \big[\wedge^n \pi
- n (\wedge^{(n-1)} \pi)\wedge(P\otimes P)\big] ,
\eeq
and the scalar dual is
\bea
\zdual{\wedge^6 M_0} 
&=& 
\frac{p^{12}}{2^6 (\ze^*\cdot\ze)^2}
\big[\ze\cdot\wedge^r \pi\cdot\ze^*
- 6\, \ze\cdot\big((\wedge^5 \pi)\wedge(P\otimes P)\big)\cdot\ze^* \big] .
\qquad
\eea
The first term in the brackets is just $\ze^*\cdot\ze$.
The second term gives
$-\ze^*\cdot\ze P\cdot P = -\ze^*\cdot\ze \tr P = -3\ze^*\cdot\ze$.
The four gauge vectors can be written as 
$(z^\ka)^\mn = (Z^\ka)^{(\mu} p^{\nu)}$,
$\ka = 1,\ldots,4$,
where $(Z^\ka)^\mu$ are four independent vectors.
This implies
$\ze^*\cdot\ze = 6!\, 2^5\, |Z|^2 p^8$, 
where $Z = \det(Z^\ka{}_\al)$.
Putting together the pieces yields 
\beq
\zdual{\wedge^6 M_0}
= -\frac{p^4}{6!\,2^{10} |Z|^2} .
\eeq
The dispersion relation for linearized GR 
is thus found to be $p^4=0$,
matching the standard result.

Next,
consider the case where linearized GR
is corrected by generic gauge-invariant terms.
This case has been explicitly treated in Ref.\ \cite{km16}.
We write $M=M_0+\de M$,
where $M_0$ is given by Eq.\ \rf{GRM}
and $\de M$ has at least the gauge invariances of $M_0$ 
but is otherwise arbitrary.
We can simplify this case by noting that 
the factor $\pi-P\otimes P$ of projection operators 
appearing in $M_0$
obeys $(\pi-P\otimes P)\cdot (\pi-\half P\otimes P) = \pi$.
The combination $\TR = \pi-\half P\otimes P$
can therefore be viewed as the gauge-invariant inverse
of $\pi-P\otimes P$,
which suggests defining
\bea
\MB &=& M\cdot\TR
= \big(\half p^2 \pi +\de\MB \big),
\notag\\
\de\MB 
&=& \de M \cdot \TR
= \de M \cdot (I - \half \et \otimes \et)\cdot\pi,
\eea
where $I-\half \et\otimes\et$ is the trace-reversal operator.
A short derivation then reveals that 
\beq
\zdual{\wedge^6 M}
= \zdual{\big(\wedge^6 \MB\cdot \wedge^6(\pi - P\otimes P)}\big)
= -2 \zdual{\wedge^6 \MB} ,
\eeq
so the dispersion relation for $M$
is the same as that for $\MB$.
Direct calculation gives
\beq
\zdual{\wedge^6 M}
= - \frac{1}{6!2^{10} |Z|^2}
\sum_{n=0}^{6}
2^{n} p^{4-2n}
\tr(\wedge^{n}\de\MB).
\label{gaugeinv}
\eeq
With the reasonable assumption that
higher-order terms remain finite as $p^2\to 0$,
this implies the leading-order covariant dispersion relation
\beq
p^4
+ 2 p^2 \de\MB_1
+ 2 (\de\MB_1^2 - \de\MB_2) =0,
\eeq
where $\de\MB_n = \tr(\de\MB^n)$.
The solution for the two perturbative modes is
\beq
p^2 = -\de\MB_1 \mp \sqrt{2\de\MB_2 - \de\MB_1^2} .
\label{gidr}
\eeq
Upon explicit evaluation of the traces
for the general gauge-invariant terms listed in Table \ref{table1},
this equation reduces correctly to the results (5) and (6) 
for the dispersion relation given in Ref.\ \cite{km16}.

With these previously known examples reproduced,
we turn to the case where linearized GR 
is instead corrected by arbitrary gauge-violating terms. 
The covariant dispersion relation takes the form \rf{drfull}
with $\cS\supseteq \cS'$,
so to evaluate it explicitly we must determine 
the modified duals that appear in Eq.\ \rf{dr2}.
They are
\beq
\zdual{\wedge^n M_0}
= \frac{p^{2n}}{2^n} \zdual{\big(\wedge^n \pi
- n (\wedge^{(n-1)} \pi)\wedge(P\otimes P)\big)} .
\eeq
After some calculation,
we find the covariant dispersion relation to be
\beq
\sum_n \frac{(n+\sB)!}{n!}
2^np^{4-2n}
\tr\left[
\big((\wedge^n\TR)\wedge\xi\big)\cdot(\wedge^{(n+\sB)}\de M)
\right] =0.
\label{dr4}
\eeq
The number of terms in the sum 
is restricted by the rank $\de r$ of $\de M$,
which may be less than the total rank $\rU$.
The limits on the sum are thus $0 \leq n\leq \de r - \sB$.

The result \rf{dr4} is the exact covariant dispersion relation
for any linearized model of gravity,
with or without gauge violation 
and with or without Lorentz violation.
For example,
the gauge-invariant scenario \rf{gaugeinv}
is contained as the special case when $\rU =6$.
The framework developed here therefore realizes
the desired goal of a model-independent approach 
to modifications of linearized gravity.
It provides calculational tools for arbitrary models
and also permits identifying generic features.
As an example of the latter,
we can see that the complexity of the exact covariant dispersion relation
is determined by the rank $\de r$ of $\de M$
and the number $\sB$ of symmetries of $M_0$ 
broken by $\de M$.
In particular,
the number of terms in the exact dispersion polynomial is $\de r-\sB+1$,
which can constrain physical aspects of the solutions.
Consider,
for instance, 
the case where rank of $\de M$
matches the number of gauge symmetries broken by $\de M$,
$\de r =\sB$.
Only one term then survives in the dispersion relation.
Since that term is proportional to $p^4$,
we find the striking result that all models of this type
must leave unaffected the conventional dispersion relation $p^2=0$.

As an explicit illustration,
consider any model in which 
only a single gauge symmetry associated with a vector $\zB$ is broken.
This implies $\rU = 7$ and $\xi=\zB\otimes \zB^*$,
and the first few terms 
of the exact covariant dispersion relation are calculated to be 
\bea
0&=&
p^4 \zB^*\de M \zB
+ 2 p^2 \left[\tr(\de\MB) \zB^*\de M \zB - \zB^*\de\MB\de M \zB \right]
\notag\\&&
+ 2 \left[\tr(\de\MB)^2 -\tr(\de\MB\de\MB)\right] \zB^*\de M \zB
\notag\\&&
+ 4 \zB^*\de\MB\de\MB\de M \zB
- 4 \tr(\de\MB)\, \zB^*\de\MB\de M \zB
+ \ldots
\qquad
\label{example}
\eea
When the rank $\de r$ of $\de M$ is one, 
this result reduces to the monomial $p^4 \zB^*\de M \zB = 0$
and so gives the usual dispersion relation $p^2=0$,
in agreement with the conclusion above.
However,
when $\de r >1$,
the dispersion relation becomes a polynomial,
and the solutions can describe modifications 
to the behavior of the gravitational modes.

Since the covariant dispersion relation \rf{dr4} is exact,
it governs all gravitational modes.
However,
inspection reveals that terms 
having larger values of $n$ involve traces
of higher powers of $\de M$,
so Eq.\ \rf{dr4} is naturally configured 
for perturbative reasoning concerning the usual modes of GR.
Although the terms with $n\geq 3$ 
naively contain inverse powers of $p^2$,
in fact this behavior is excluded by continuity
and so any contributions from them are perturbatively small. 
The three terms with $n=0,1,2$ 
are therefore the ones of perturbative relevance,
and they yield a quadratic dispersion relation
whose solution describes perturbative effects 
on the usual gravitational modes of GR.
This structure can be seen explicitly in the example \rf{example}
with only a single broken gauge symmetry.
The solution in this case takes the generic form \rf{gidr},
with the factors of  $\de\MB_1$ and $\de\MB_2$
replaced by factors of the polyomial coefficients 
in Eq.\ \rf{example}.

Another interesting special case is the set of models
with coefficients having only purely temporal components,
which is a subset of the isotropic limit.
All gauge-invariant models of this type 
are discussed in  Ref.\ \cite{km16}.
For example,
the operator $\cd{d}{}^{\mu\rh\dc\nu\si\dc\dc^{d-4}}$
contains models with purely temporal components
that can be isolated by working with its double dual 
and restricting attention to the components $(\std{d})^{00\ldots}$
with $d-2$ temporal indices.
Observational constraints on these coefficients 
have been placed using gravitational waves and other techniques
\cite{2013Bailey,2014Shao,le16,kt15,jt16,yu16,2017Abbott,km17}.

Many of the gauge-violating representations
listed in Table \ref{table1} 
also contain coefficients with purely temporal components.
Typically,
these generate nontrivial effects on the gravitational modes.
The observational implications of these 
lie beyond our present scope
and offer an interesting open direction for future investigation.
Note,
however,
that some of these cases may produce no measurable effects.
Consider,
for example,
the operator $\Kzz{}^{\mn\rs\dc^{d-2}}$ 
restricted to the components $\Kzz{}^{00\ldots}$ 
with $d+2$ temporal indices.
The matrix $\de M$ is then of rank $\de r =1$
and has nonzero component
$\de M_{00}{}^{00} = \sum_d \Kzz{}^{00\ldots} E^{d-2}$.
A single broken gauge vector exists,
which can be taken as $\zB^\mn = \et^{0(\mu} p^{\nu)}$,
so $\sB=1$ as well.
The dispersion relation becomes 
$p^4 \sum_d \Kzz{}^{00\ldots} E^{d}=0$
and reduces to $p^2=0$,
in agreement with the general result for $\de r =\sB$ discussed above.
These coefficients therefore have no effect 
on the behavior of the usual gravitational modes.

More involved cases exist 
that also leave unaffected the usual gravitational modes.
One example with $\de r = \sB = 2$
involves the CPT-odd operator 
$\Qaa{}^{\mu\rh\dc\nu\si\dc^{d-3}}$ for $d=3$.
Taking the dual of this and restricting
to the purely temporal coefficient
$\widetilde{\Q}^{(3,3)}{}^{000}$ 
produces a rank-two matrix $\de M$ with nonvanishing components
$\de M^{0j0k} = - i \widetilde{\Q}^{(3,3)}{}^{000} \ep_{jkl} p^l/8$
that is symmetric under interchange of the first or second pair of indices
and is antisymmetric under interchange of the pairs.
Two unbroken gauge vectors exist,
which can be taken as
$(\zU^0)^\mn = \et^{0(\mu} p^{\nu)}$ and
$(\zU^1)^\mn = p^\mu p^\nu$.
Calculation shows the dispersion relation is 
$ p^4 (\widetilde{\Q}^{(3,3)}{}^{000})^2 E^4 = 0$
and so again yields $p^2=0$,
as expected.
Other components of ${\Q}^{(3,3)}{}^{\mu\rh\la\nu\si}$ can,
however,
modify gravitational propagation
\cite{fe07}.

To summarize,
we have provided in this work 
a framework for studying diffeomorphism and Lorentz violations
in linearized gravity theories.
The techniques developed here yield the classification and enumeration 
of all gauge-invariant and gauge-violating terms
in the general effective field theory for the metric fluctuation.
For the various possible scenarios,
we have obtained the exact covariant dispersion relations
for the gravitational modes.
The expressions hold for operators of arbitrary mass dimension,
and reduce to known results in suitable special limits.
Results for any specific model of linearized gravity
can be extracted as a special limit.
The work opens the path to model-independent phenomenological studies 
of arbitrary gauge and Lorentz violation in nature,
representing a broad arena for search and discovery.

This work was supported in part 
by the United States Department of Energy
under grant number {DE}-SC0010120,
by the United States National Science Foundation 
under grant number PHY-1520570,
and by the Indiana University Center for Spacetime Symmetries.

\end{document}